\documentclass[%
prl,
,twocolumn%
,superscriptaddress
,amsmath,amssymb,aps,nobibnotes,
nofootinbib
,preprintnumbers
]{revtex4-1}
\usepackage[pdftex]{graphicx}
\usepackage{bm}
\usepackage{dcolumn}
\usepackage{color}
\usepackage{tabularx}
\def\urm#1{\scriptstyle{\text{\textrm{\textmd{\textup{#1}}}}}}

\def\rs{r_{\urm{s}}}
\def\ve#1{{\bm{#1}}}
\def\ec{\epsilon_{\urm{corr}}}
\def\ecgb{\epsilon_{\urm{corr}}^{\urm{GB}}}

\def\decref{\Delta \epsilon_{\urm{corr,ref}}}
\def\decfit{\Delta \epsilon_{\urm{corr,fit}}}
\def\defeq{\mathrel{:=}}

\let\temp\epsilon
\let\epsilon\varepsilon
\let\varepsilon\temp
\let\temp\relax
\begin{document}
\title{\textit{Ab initio} construction of the energy density functional 
  for electron systems
  with the functional-renormalization-group-aided density functional theory}
\author{Takeru Yokota}
\email{tyokota@issp.u-tokyo.ac.jp}
\affiliation{
  Institute for Solid State Physics, The University of Tokyo, Kashiwa, Chiba 277-8581, Japan
}
\author{Tomoya Naito}
\email{tomoya.naito@riken.jp}
\affiliation{
  Department of Physics, Graduate School of Science, The University of Tokyo, Tokyo 113-0033, Japan}
\affiliation{
  RIKEN Nishina Center, Wako 351-0198, Japan}
\date{\today}
\preprint{RIKEN-QHP-482}
\preprint{RIKEN-iTHEMS-Report-20}
\begin{abstract}
  We show an \textit{ab initio} construction of the energy density functional (EDF) for electron systems using the functional renormalization group.
  The correlation energies of the homogeneous electron gas given in our framework 
  reproduce the exact behavior at high density and
  agree with the Monte-Carlo data in a wide range of densities.
  Our analytic technique enables us to get the correlation energies efficiently for various densities,
  which realizes the determination of EDF in the local density approximation (LDA) without any fitting for physically relevant densities.
  Applied to the Kohn-Sham calculation for the noble gas atoms, our EDF shows comparable results to those of other conventional ones in LDA.
\end{abstract}
\maketitle
\textit{Introduction\/.}~Density functional theory (DFT) \cite{hoh64} is 
a successful framework to analyze quantum many-body systems 
providing an efficient way
known as the Kohn-Sham (KS) scheme \cite{koh65} 
and has been employed in various fields,
including condensed matter physics, quantum chemistry, and nuclear physics.
DFT is often positioned as a first-principles method.
However, most of the energy density functionals (EDFs), 
which govern the accuracy of DFT calculations, are empirically constructed,
and the recipe to systematically construct EDF 
based on microscopic Hamiltonians has not been established yet \cite{per01,med17,kep17}.
\par
In this Letter,
we focus on an attempt for the microscopic construction of the EDF
put forward in Refs.~\cite{pol02,sch04},
which we call the functional-renormalization-group-aided DFT (FRG-DFT).
This is based on the functional renormalization group (FRG) \cite{weg73,wil74,pol84,wet93}
(for reviews, see, e.g., Refs.~\cite{ber02,paw07,gie12,met12,dup20}),
which is an established method for quantum many-body systems:
In FRG, one-parameter exact flow equation for the effective action,
the quantum counterpart of the classical action,
is utilized to non-perturbatively include quantum or thermal fluctuations.
Owing to the fact that the EDF can be defined by 
the effective action $ \Gamma \left[ \rho \right] $ with the local density $ \rho $ \cite{fuk94,fuk95,val97},
accumulated methods in FRG are expected to be applied to the construction of EDF.
\par
Applications of the FRG-DFT accomplished recently include analysis of the ground states 
in lower-than-$ \left( 1+1 \right) $-dimensional systems \cite{kem13,kem17a,lia18,yok18}
and $ \left( 2+1 \right) $-dimensional homogeneous electron gas \cite{yok19},
excited states of a $ \left( 1+1 \right) $-dimensional systems \cite{yok18b},
and formalism for superfluid systems \cite{yok20}.
However, there has been no numerical application to the $ \left( 3+1 \right) $-dimensional systems yet,
which must be achieved to establish the FRG-DFT as a practical method.
In particular, as systems for which DFT is frequently employed,
the electron systems are one of the most important targets.
\par
The aim of this Letter is the microscopic derivation of the EDF
$ E \left[ \rho \right] $ for the spin-unpolarized electron systems 
with the aid of the FRG-DFT.
To our best knowledge, 
this work is the first numerical application of the FRG-DFT to $ \left( 3+1 \right) $-dimensional systems.
As a first step of the microscopic construction of the EDF,
we consider the exchange-correlation part 
in the local density approximation (LDA)
and aim at the construction of the correlation part.
To this end,
we apply the FRG-DFT to the $ \left( 3+1 \right) $-dimensional homogeneous electron gas (3DHEG),
derive the expression for the correlation energy per particle $ \ec $
by solving the flow equation analytically with employing the second-order vertex expansion,
and obtain $ \ec \left( \rs \right) $ as a function of the Wigner-Seitz radius
$ \rs = \left[ 3 / \left( 4 \pi \rho \right) \right]^{1/3} $.
Our $ \ec \left( \rs \right) $ reproduces the exact behavior at the high-density limit
given by the result of Gell-Mann--Brueckner resummation $ \ecgb \left( \rs \right) $ \cite{gel57}
and agrees with the results of the diffusion Monte-Carlo (DMC) calculations \cite{cep80,zon02,spi13} in a wide range of densities.
\par
A concern about the application to $ \left( 3+1 \right) $-dimensional systems
may be that the coordinate or momentum integrals 
in the FRG-DFT calculation
become time consuming.
For 3DHEG, however, we find that the dimension of the integrals 
can be drastically reduced with analytic techniques,
which enables us to obtain $ \ec \left( \rs \right) $
densely enough to determine the EDF without fitting for physically relevant densities.
This is in contrast to other conventional LDA EDFs,
most of which are determined based on empirical choice of the fitting function for a few DMC data points.
\par
Furthermore,
applying the KS calculation of the ground states of the noble gas atoms,
we demonstrate that the EDF constructed from our FRG-DFT data
shows comparable results to other conventional LDA EDFs.
\par
In this Letter, Hartree atomic units are employed.
\par
\textit{FRG-DFT\/.}~We briefly summarize the formalism of FRG-DFT and our analytic results in the 3DHEG,
where electrons and background ions neutralizing the system interact to each other via the two-body Coulomb interaction 
$ U \left( \ve{x} \right) = 1 / \left| \ve{x} \right| $.
Following Refs.~\cite{pol02,sch04}, we consider the evolution 
when the inter-particle interaction is gradually turned on.
Let us employ a parametrized two-body interaction
$ U_{\lambda} \left( \ve{x} \right) $ with the evolution parameter $ \lambda $
running from $ \lambda = 0 $ to $ 1 $ and 
set $ U_{\lambda = 0} \left( \ve{x} \right) = 0 $
and
$ U_{\lambda = 1} \left( \ve{x} \right) = U \left( \ve{x} \right) $
to describe the evolution from the free 
to the fully interacting systems.
Not only the electron-electron 
but also the electron-ion and ion-ion interactions 
are substituted by $ U_{\lambda} \left( \ve{x} \right) $
so as to keep the system neutral and avoid divergence 
caused by the Hartree energy during the evolution \cite{yok19}.
\par
The key quantity of the FRG-DFT is the effective action 
$ \Gamma_{\lambda} \left[ \rho \right] $ for density $ \rho $. 
To define it, we start from the action depending on $ \lambda $ in the imaginary-time formalism:
\begin{align*}
  S_{\lambda}
  \left[ \psi, \psi^{\dagger} \right]
  = & \,
      \int_{X}
      \psi^{\dagger} \left( X_\epsilon \right) 
      \left(
      \partial_\tau
      -
      \frac{\nabla^2}{2}
      \right)
      \psi \left( X \right) \\
    & \,
      +
      \frac{1}{2}
      \int_{X,X'}
      U_{\urm{2b}, \lambda}
      \left( X, X' \right)
      \hat{\rho}_{\Delta} \left( X \right)
      \hat{\rho}_{\Delta} \left( X' \right).
\end{align*}
Here, we have introduced $ X = \left( \tau, \ve{x} \right) $ and 
$ \int_X = \int d \tau \int d \ve{x} $
with imaginary time $ \tau $ and spatial coordinate $ \ve{x} $,
$ U_{\urm{2b}, \lambda} \left( X, X' \right) =
\delta \left( \tau - \tau' \right) U_{\lambda} \left( \ve{x} - \ve{x}' \right) $,
$ \psi \left( X \right)
= {}^{\urm{t}} \! \left( \psi_{\uparrow} \left( X \right), \psi_{\downarrow} \left( X \right) \right) $
standing for the electron field
with spin $ \uparrow $ and $ \downarrow $,
and $ \hat{\rho}_{\Delta} \left( X \right) = \hat{\rho} \left( X \right) - n_e $
with the density field
$ \hat{\rho} \left( X \right) = \psi^{\dagger} \left( X_{\epsilon} \right) \psi \left( X \right) $
and $ n_e = 3 / \left( 4 \pi \rs^3 \right) $
being the densities of electrons and back-ground ions.
The second term includes
the electron-electron, electron-ion, and ion-ion interaction terms.
We have also introduced $ X_{\epsilon} = \left( \tau + \epsilon, \ve{x} \right) $
with an infinitesimal $ \epsilon > 0 $ 
so that the Hamiltonian corresponding to $ S_{\lambda} $ becomes normal ordered \cite{yok18}.
Then, $ S_{\lambda} $ defines $ \Gamma_{\lambda} \left[ \rho \right] $ as
\begin{align*}
	\Gamma_{\lambda} \left[ \rho \right]
	= \sup_{J} \left( \int_X J \left( X \right) \rho \left( X \right)
	- \ln Z_{\lambda} \left[ J \right] \right),
\end{align*}
where
\begin{align*}
	Z_{\lambda} \left[ J \right]
= \int \mathcal{D} \psi \, \mathcal{D} \psi^{\dagger} \,
e^{- S_{\lambda} \left[ \psi, \psi^{\dagger} \right]
  + \int_X J \left( X \right) \hat{\rho} \left( X \right)}
\end{align*}
is the generating functional for density correlation functions
and $\rho(X)$ is an arbitrary density.
A notable feature of $ \Gamma_{\lambda} \left[ \rho \right] $
is that it satisfies the variational principle and gives
the ground-state energy and density \cite{fuk94,val97},
which means that the EDF $ E_{\lambda} \left[ \rho \right] $
is identified with $ \Gamma_{\lambda} \left[ \rho \right] $ as
$ E_{\lambda} \left[ \rho \right] = \lim_{\beta \to \infty} \Gamma_{\lambda} \left[ \rho \right] / \beta $
with the inverse temperature $ \beta = \int d \tau $.
\par
The key equation in the FRG-DFT is the evolution equation
determining $ \Gamma_{\lambda} \left[ \rho \right] $ \cite{sch04,kem13,kem17a,yok18,yok19}:
\begin{align}
  \partial_{\lambda} \Gamma_{\lambda} \left[ \rho \right]
  = & \, 
      \frac{1}{2}
      \int_{X,X'}
      \partial_{\lambda} U_{\urm{2b}, \lambda} \left( X, X' \right)
      \left[
      \vphantom{
      \Gamma_{\lambda}^{{\left( 2 \right)}-1} \left[ \rho \right] \left( X_{\epsilon'}, X' \right)
      -
      \rho \left( X \right)
      \delta \left( \ve{x} - \ve{x}' \right)
      }
      \rho_{\Delta} \left( X \right)
      \rho_{\Delta} \left( X' \right)
      \right.
      \notag \\
    & \,
      \left.
      +
      \Gamma_{\lambda}^{{\left( 2 \right)}-1} \left[ \rho \right] \left( X_{\epsilon'}, X' \right)
      -
      \rho \left( X \right)
      \delta \left( \ve{x} - \ve{x}' \right)
      \right],
      \label{eq:flow}
\end{align}
where $ \rho_{\Delta} \left( X \right) = \rho \left( X \right) - n_e $ 
and $ \Gamma^{{\left( 2 \right)}-1}_{\lambda} \left[ \rho \right] \left( X, X' \right) $
being the inverse of 
$ \frac{\delta^2 \Gamma_{\lambda} \left[ \rho \right]}{\delta \rho \left( X \right) \, \delta \rho \left( X' \right)}$, which satisfies 
\begin{align*}
	\int_{X''} \Gamma^{{\left( 2 \right)}-1}_{\lambda} \left[ \rho \right] \left( X, X'' \right)
\frac{\delta^2 \Gamma_{\lambda} \left[ \rho \right]}{\delta \rho \left( X'' \right) \, \delta \rho \left( X' \right)}
= \delta \left( X-X' \right).
\end{align*}
Also, $ X_{\epsilon'} $ is defined in the same manner as $ X_{\epsilon} $
but $ \epsilon' \to 0 $ limit is taken after $ \epsilon \to 0 $ so that 
$ \Gamma_{\lambda}^{{\left( 2 \right)}-1} \left[ \rho \right] \left( X_{\epsilon'}, X' \right) $
can be treated as the density correlation function \cite{yok18}.
The crucial point of Eq.~\eqref{eq:flow} is that it is written in
a closed form of $ \Gamma_{\lambda} \left[ \rho \right] $,
which provides systematic schemes for the derivation of $ \Gamma_{\lambda} \left[ \rho \right] $.
\par
Practically, the functional differential equation \eqref{eq:flow} needs to be converted to some numerically solvable equations.
Here, we introduce the vertex expansion \cite{pol02,sch04}:
The functional Taylor expansion around a homogeneous density $ \rho \left( X \right) = n_e $
is applied to Eq.~\eqref{eq:flow}, which yields a hierarchy of 
differential equations for density correlation functions \cite{kem17a,yok18,yok18b,yok19}.
We consider the expansion up to the second order and truncate higher-order terms.
The equations up to the second order in the momentum-space representation read
\begin{align}
  \partial_{\lambda} \epsilon_{\urm{gs}, \lambda}
  & = 
    \frac{1}{2 n_e}
    \int_{\ve{p}}
    \partial_{\lambda} \tilde{U}_{\lambda} \left( \ve{p} \right) 
    \left[
    \int_\omega
    e^{i \omega \epsilon'}
    \tilde{G}^{\left( 2 \right)}_{\lambda} \left( P \right)
    -
    n_e
    \right],
    \label{eq:fene} \\
  \partial_{\lambda} \tilde{G}_{\lambda}^{\left( 2 \right)} \left( P \right)
  & = 
    -
    \partial_{\lambda}
    \tilde{U}_{\lambda} \left( \ve{p} \right) 
    \left[
    \tilde{G}_{\lambda}^{\left( 2 \right)} \left( P \right)
    \right]^2
    +
    C_{\lambda} \left( P \right),
    \label{eq:fg2t}
\end{align}
where $\tilde{U}_\lambda({\bm p})$ is the Fourier transform of $U_\lambda({\bm x})$
and 
$ \epsilon_{\urm{gs}, \lambda}
= \lim_{\beta \to \infty} \Gamma_{\lambda} \left[ n_e \right] / \left( \beta N \right) $
with $ N = n_e \int d \ve{x} $ being the total particle number
is the ground-state energy per particle.
Here, we have introduced $ P = \left( \omega, \ve{p} \right) $ and 
$ \int_{P} = \int_\omega \int_{\ve{p}}
= \int d \omega/ \left( 2 \pi \right) \int d \ve{p}/ \left( 2 \pi \right)^3 $
with the Matsubara frequency $ \omega $ and the spatial momentum $ \ve{p} $,
\begin{align}
  C_{\lambda} \left( P \right)
  = & \,
      - \frac{1}{2}
      \int_{P'}
      \partial_{\lambda}
      \tilde{U}_{\lambda} \left( \ve{p}' \right)
      \left[
      \tilde{G}^{\left( 4 \right)}_{\lambda} \left( P', -P', P \right)
      \right.
      \notag \\
    & \,
      -
      \left.
      \tilde{G}_{\lambda}^{\left( 2 \right)} \left( 0 \right)^{-1}
      \tilde{G}_{\lambda}^{\left( 3 \right)} \left( P',-P' \right)
      \tilde{G}_{\lambda}^{\left( 3 \right)} \left( P, -P \right)
      \right],
      \label{eq:c0}
\end{align}
and $ \tilde{G}_{\lambda}^{\left( n \right)} \left( P_1, \ldots, P_{n-1} \right) $
being the connected density correlation function.
Since $ C_{\lambda} \left( P \right) $ is composed of higher-order correlation functions,
an approximation for it is required.
Here, we employ the approximation
$ C_{\lambda} \left( P \right) \approx C_{\lambda=0} \left( P \right) $,
with which Eqs.~\eqref{eq:fene} and \eqref{eq:fg2t} can be solve analytically
when $ U_{\lambda} \left( \ve{x} \right) $ is chosen as 
$ U_{\lambda} \left( \ve{x} \right) = \lambda U_{\lambda=1} \left( \ve{x} \right) $ \cite{yok19}.
\par
Extracting $ \ec $ from $ \epsilon_{\urm{gs}, \lambda=1} $, we obtain
\begin{equation}
  \label{eq:esol}
  \ec \left( \rs \right)
  =
  \frac{1}{2 n_e}
  \int_{P}
  \left[
    \ln f \left( A_P, B_P \right)
    -
    A_P
  \right],
\end{equation}
which plays the central role in our construction of the EDF.
Here, we have introduced
$ f \left( x, y \right) = \cosh y + \left( x / y \right) \sinh y $, 
$ A_P \defeq \tilde{U} \left( \ve{p} \right) \tilde{G}_{\lambda=0}^{\left( 2 \right)} \left( P \right) $,
and
$ B_P \defeq \left[ \tilde{U} \left( \ve{p} \right) C_{\lambda=0} \left( P \right) \right]^{1/2}$,
in which 
$\tilde{U}({\bm p})=4\pi / {\bm p}^2$ is the Coulomb interaction in the momentum representation and
$ C_{\lambda=0} \left( P \right) $ is evaluated from
the connected density
correlation function in the free system:
\begin{align*}
	&\tilde{G}^{\left( n \right)}_{\lambda=0} \left( P_1, \ldots, P_{n-1} \right)
	\\
	&= 
	- 2 \sum_{\sigma \in S_{n-1}}
	\int_{P'} \prod_{k = 0}^{n-1}
	\tilde{G}_{\urm{F}, 0}^{\left( 2 \right)} \left( \sum_{i = 1}^{k} P_{\sigma(i)} + P' \right) 
\end{align*}
with the symmetric group $ S_{n-1} $ of order $ n-1 $,
the two-point propagator of free fermions
$ \tilde{G}_{\urm{F}, 0}^{\left( 2 \right)} \left( P \right)
= \left[ i \omega - \xi \left( \ve{p} \right) \right]^{-1} $,
$ \xi \left( \ve{p} \right) \defeq \ve{p}^2/2 - p_{\urm{F}}^2/2 $,
and the Fermi momentum  $ p_{\urm{F}} = \left( 9 \pi / 4 \right)^{1/3} / \rs $.
\par
Before ending the summary of the formalism, 
we comment on the behavior of Eq.~\eqref{eq:esol} at the high-density limit $ \rs \to 0 $: 
Through the use of the scaling behavior 
\begin{align*}
	\left. \tilde{G}_{\lambda=0}^{\left( 2 \right)} \left( P \right) \right|_{\rs}
=&
\left.
  \rs^{-1} 
  \tilde{G}_{\lambda=0}^{\left( 2 \right)} \left( \overline{P} \right)
\right|_{\rs = 1},
\\
\left. C_{\lambda=0} \left( P \right) \right|_{\rs}
=&
\left.
  C_{\lambda=0} \left( \overline{P} \right)
\right|_{\rs = 1},
\end{align*}
with the dimensionless momentum $ \overline{P} = \left( \rs^2 \omega, \rs \ve{p} \right) $,
the expansion of Eq.~\eqref{eq:esol} with respect to $ \rs $ is obtained.
From the expansion, one finds that the exact behavior 
$ \ecgb \left( \rs \right) $ is reproduced at $ \rs \to 0 $.
\par
\textit{Reduction of dimension of multi-integral\/.}~A difficulty in the three-dimensional system may be that the numerical 
evaluation of
multi-integrals 
with respect to momenta is too costly to get $ \ec \left( \rs \right) $ for various $ \rs $.
We find that this can be circumvented since the dimension of the 
integral
in Eq.~\eqref{eq:c0} can be drastically reduced in an analytic manner.
\par
Using the expression for
$ \tilde{G}^{\left( n \right)}_{\lambda=0} \left( P_1, \ldots, P_{n-1} \right) $
given in the sentence below Eq.~\eqref{eq:esol}
and performing the frequency integral, Eq.~\eqref{eq:c0} becomes
\begin{align}
  C_{\lambda=0} \left( P \right)
  = & \,
      2
      \sum_{s=0,1}
      \left( -1 \right)^{s+1}
      \int_{\ve{p}', \ve{p}''}
      \tilde{U} \left( \ve{p}' - \ve{p}'' - s \ve{p} \right)
      \theta_{\ve{p}'}
      \theta_{\ve{p}''}
      \notag \\
    & \,
      \times
      \left[
      D \left( \omega, \ve{p}, \ve{p}' - s \ve{p} \right)
      -
      D \left( \omega, \ve{p}, \ve{p}'' \right)
      \right]^2,
      \label{eq:c_freq}
\end{align}
where
\begin{align*}
	D \left( \omega, \ve{p}, \ve{p}' \right)
=
\left[ i \omega - \xi \left( \ve{p}' + \ve{p} \right) + \xi \left( \ve{p}' \right) \right]^{-1},
\end{align*}
and $ \theta_{\ve{p}} = \theta \left( - \xi \left( \ve{p} \right) \right) $ 
with the Heaviside step function $ \theta \left( x \right) $.
By using
$ \tilde{U} \left( \ve{p} \right) = \int d \ve{x} \, e^{i \ve{p} \cdot \ve{x}}/ \left| \ve{x} \right| $
and employing the cylindrical coordinates
with choosing the direction of $ \ve{p} $ as the direction of the longitudinal axis ($ z $ axis),
Eq.~\eqref{eq:c_freq} is rewritten as follows:
\begin{align}
  \label{eq:cf}
  & C_{\lambda=0} \left( P \right)
    \notag \\
  = & \,
      \sum_{s=0,1}
      \frac{\left( -1 \right)^{s+1}}{2 \pi^3}
      \int_{-p_{\urm{F}}}^{p_{\urm{F}}}
      dp'_z
      \int_{-p_{\urm{F}}}^{p_{\urm{F}}}
      dp''_z \,
      P_{r} \left( p_z' \right)
      P_{r} \left( p_z'' \right)
      \notag \\
  & \,
    \times
    \left[
    D' \left( \omega, p, p_z' \right)
    -
    D' \left( \omega, p, p_z''-sp \right)
    \right]^2
    \notag \\
  & \,
    \times
    I \left(
    P_{r} \left(p_z' \right),
    P_{r} \left(p_z'' \right),
    \left| p_z' - p_z'' + s p \right|
    \right).
\end{align}
Here, we have introduced 
$ P_r \left( p_z \right) = \left( p_{\urm{F}}^2 - p_z^2 \right)^{1/2} $,
$ D' \left( \omega, p, p_z' \right) = \left( i \omega + pp_{z}' + p^2/2 \right)^{-1} $,
and
\begin{equation*}
  I \left( a, b, c \right)
  =
  \int_{0}^{\infty}
  dr \,
  \frac{1}{r}
  J_1 \left( ar \right)
  J_1 \left( br \right)
  K_0 \left( cr \right),
\end{equation*}
where $ J_1 \left( x \right) $ and $ K_0 \left( x \right) $ are the Bessel function of the first kind
and the modified Bessel function of the second kind, respectively.
Then, the integral in $ I \left( a, b, c \right) $ can be performed 
analytically \cite{zwi14}:
\begin{align}
  I \left( a, b, c \right)
  = & \,
      l_{1}^2 \left( a, b, c \right)
      +
      b^2
      \ln \left(
      1
      -
      \frac{l_1^2 \left( a, b, c \right)}{b^2}
      \right) 
      \notag \\
    & \,
      +
      c^2
      \ln \left(
      1
      -
      \frac{l_1^2 \left( a, b, c \right)}{c^2}
      \right),
      \label{eq:I}
\end{align}
where
\begin{align*}
	l_1 \left( a, b, c \right)
	= \frac{1}{2}\left[ \sqrt{\left( b + c \right)^2 + a^2} - \sqrt{\left( b - c \right)^2 + a^2} \right].
\end{align*}
Finally, Eq.~\eqref{eq:cf} together with Eq.~\eqref{eq:I} shows that
only a double integral is required for the calculation of 
$ C_{\lambda=0} \left( P \right) $.
Needless to say, the isotropy reduces the dimension of the integral in Eq.~\eqref{eq:esol}.
\par
\textit{Results of the correlation energy\/.}~The reduction of the dimension of the integral saves the time for 
the numerical calculation and enables one to obtain $ \ec $ for many $ \rs $.
The calculation was carried out on 65536 grid points with the logarithmic mesh
in $ \rs \in \left[ 10^{-6} \, \mathrm{a.u.} , 100 \, \mathrm{a.u.} \right) $.
\par
\begin{figure}[!t]
  \centering
  \includegraphics[width=\columnwidth]{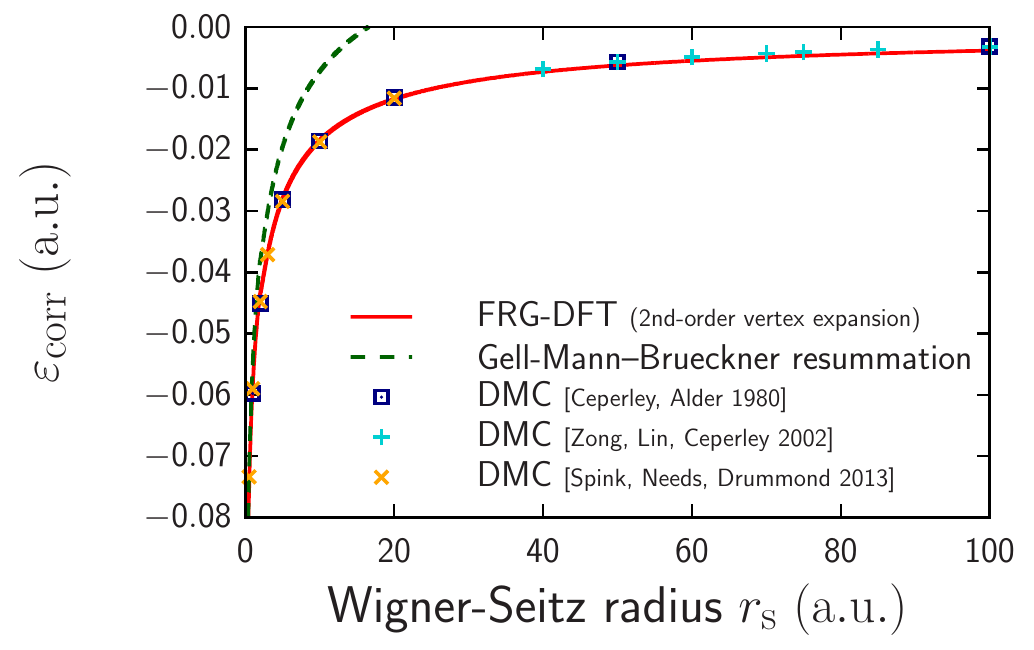}
  \caption{
    Correlation energy par particle, $ \ec $, of the 3DHEG 
    derived by the FRG-DFT method as a function of 
    $ \rs $.
    The results derived by the Gell-Mann--Brueckner resummation
    and the diffusion Monte-Carlo (DMC) calculations are also shown.
    The DMC results are obtained by
    subtracting the kinetic and exchange energies
    \cite{sla51,mah00} 
    from the total energies 
    given in Refs.~\cite{cep80,zon02,spi13}.}
  \label{fig:corr}
\end{figure}
Figure~\ref{fig:corr} shows $ \ec \left( \rs \right) $ obtained by the FRG-DFT
together with $ \ecgb \left( \rs \right) $
and the DMC results
\cite{cep80,zon02,spi13}.
As expected from our analytical discussion, one
can see the FRG-DFT result
reproduces the exact behavior
at the high-density limit given by $ \ecgb \left( \rs \right) $.
The FRG-DFT result 
is also consistent with the DMC results
in a wide range of $ \rs $, and in particular the agreement 
becomes better as the density increases:
The deviations from the results in Ref.~\cite{cep80}
are about $ 2.0 \, \% $ at $ \rs = 1 \, \mathrm{a.u.} $,
$ 6.7 \, \% $ at $ \rs = 50 \, \mathrm{a.u.} $,
and 
$ 17 \, \% $ at $ \rs = 100 \, \mathrm{a.u.} $
\par
Now we have shown that 
$ \ec \left( \rs \right) $ can be obtained in the framework of the FRG-DFT and becomes accurate as the density increases, which is one of the main results of this Letter.
In the remaining part, we attempt a construction of the EDF by use of our $ \ec \left( \rs \right) $.
\par
\textit{Construction of the energy density functional\/.}~Since $ \ec $ are
obtained very densely for various $ \rs $ in our scheme,
we can construct the LDA EDF 
$ E^{\urm{LDA}}_{\urm{corr}} \left[ \rho \right]
=
\int
d \ve{x} \,
\rho \left( \ve{x} \right)
\ec \left( \rs \left( \rho \left( \ve{x} \right) \right) \right) $
without any fitting for physically relevant densities.
This is in sharp contrast 
to other conventional EDFs such as 
VWN \cite{vos80}, PZ81 \cite{per81}, and PW92 \cite{per92}, 
which are determined by fitting the few DMC data obtained by Ceperley and Alder
\cite{cep80}.
\par
Our functional, which is referred to as the FRG-numerical-table functional (FRG-NT),
is constructed as follows:
In $ \rs \in \left[ 10^{-6} \, \mathrm{a.u.}, 100 \, \mathrm{a.u.} \right) $,
$ \ec \left( \rs \right) $ are determined by the interpolation of the FRG-DFT data. 
For simplicity, we employ the linear interpolation; the results hardly depend on the choice of the interpolation function.
In $ \rs < 10^{-6} \, \mathrm{a.u.} $,
$ \ec \left( \rs \right) $ is substituted
by $ \ecgb \left( \rs \right) $.
The FRG-DFT data
are extrapolated to $ \rs \geq 100 \, \mathrm{a.u.} $ by a fitting function
$ \ec \left( \rs \right) = \gamma / \left( 1 + \beta_1 \sqrt{\rs} + \beta_2 \rs \right) $ \cite{per81}.
The fitting parameters are chosen to be 
$ \gamma = 0.0378052 $, 
$ \beta_1 = -0.801035 $,
and 
$ \beta_2 = -0.0306778 $, 
which are obtained by fitting the data in 
$ 95 \, \mathrm{a.u.} < \rs < 100 \, \mathrm{a.u.} $
\par
A remark is in order here: 
The evaluation of 
$ \ec' \left( \rs \right) = d \ec \left( \rs \right) / d \rs $ appearing in the KS potential
$ \delta E_{\urm{corr}}^{\urm{LDA}} \left[ \rho \right] / \delta \rho \left( \ve{x} \right)
=
\left[ \ec \left( \rs \right) - \left( \rs/3 \right) \ec' \left( \rs \right) \right]_{\rs = \rs \left( \rho \left( \ve{x} \right) \right)} $
with
the numerical differentiation may 
cause numerical errors.
We evade this by performing the
analytic differentiation of Eq.~\eqref{eq:esol}:
\begin{equation}
  \ec'
  \left( \rs \right)
  =
  \frac{1}{2 n_e \rs}
  \int_{P}
  \left[
    g \left( A_P, B_P \right)
    -
    2 \ln
    f \left( A_P, B_P \right)
  \right],
\end{equation}
where $ g \left( x, y \right) = x + \left( x \cosh y + y \sinh y \right) / f \left( x, y \right) $.
To perform the differentiation, we have used
$ \left. \tilde{G}_{\lambda=0}^{\left( 2 \right)} \left( P \right) \right|_{\rs}
=
\left.
  \rs^{-1}
  \tilde{G}_{\lambda=0}^{\left( 2 \right)} \left( \overline{P} \right)
\right|_{\rs = 1} $,
$ \left. C_{\lambda=0} \left( P \right) \right|_{\rs}
=
\left.
  C_{\lambda=0} \left( \overline{P} \right)
\right|_{\rs = 1} $
and rewritten Eq.~\eqref{eq:esol} in terms of $ \rs $ and quantities independent of $ \rs $.
We 
calculate $ \ec' \left( \rs \right) $
on the same grid for $ \rs $ as $ \ec \left( \rs \right) $
and determine $ \ec' \left( \rs \right) $ for arbitrary $ \rs $
in the same manner as $ \ec \left( \rs \right) $.
\par
Additionally, we prepare a functional, which we name FRG-PZ, 
by fitting the FRG-DFT data
with the same function as PZ81 
\begin{equation}
  \label{eq:PZ81}
  \ec \left( \rs \right)
  =
  \begin{cases}
    A \ln \rs + B + C \rs \ln \rs + D \rs
    & \rs < 1 \, \mathrm{a.u.}, \\
    \gamma / \left( 1 + \beta_1 \sqrt{\rs} + \beta_2 \rs \right)
    & \rs \ge 1 \, \mathrm{a.u.},
  \end{cases}
\end{equation}
for the purpose of comparing our functionals to PZ81 and discussing 
the origin of the deviation between EDFs.
Here, $ A = 0.0311 $ and $ B = -0.0480 $ reproduce
$ \ecgb \left( \rs \right) $ at $ \rs \to 0 $ 
.
The remaining parameters $ C $, $ D $, $ \gamma $, $ \beta_1 $, and $ \beta_2 $ are 
related to each other through the continuum conditions 
for $ \ec \left( \rs \right) $ and $ \ec' \left( \rs \right) $
at $ \rs = 1 \, \mathrm{a.u} $:
\begin{subequations}
  \begin{align}
    \gamma
    & =
      \left( 1 + \beta_1 + \beta_2 \right)
      \left( B + D \right),
      \label{eq:C1} \\
    \beta_2
    & = 
      -
      \frac{2 \left( A + C \right) \left( 1 + \beta_1 \right)
      + B \beta_1 + 2D + 3 \beta_1 D}
      {2 \left( A + B + C + 2D \right)}.
      \label{eq:C2}
  \end{align}
\end{subequations}
Table \ref{tab:param} lists the values of the parameters obtained by the fitting
with the conditions Eqs.~\eqref{eq:C1} and \eqref{eq:C2}.
\begin{table}[tb]
  \centering
  \caption{
    Parameters for FRG-PZ.
    For comparison, the parameters of PZ81
	\cite{per81} 
	are also shown.
    All the data are shown in the Hartree atomic units.}
  \label{tab:param}
  \begin{ruledtabular}
    \begin{tabular}{ldd}
      & \multicolumn{1}{c}{PZ81 \cite{per81}} & \multicolumn{1}{c}{FRG-PZ} \\
      \hline 
      $ C $       &  0.0020 &  0.00173055 \\
      $ D $       & -0.0116 & -0.0100569  \\
      $ \gamma $  & -0.1423 & -0.175617   \\
      $ \beta_1 $ &  1.0529 &  1.67669    \\
      $ \beta_2 $ &  0.3334 &  0.348219   \\    
    \end{tabular}
  \end{ruledtabular}
\end{table}
\par
\textit{Benchmark test of the functionals\/.}~We 
apply our EDF to the KS calculation of 
the ground-state energies of noble gas atoms and 
compare with
other conventional EDFs such as VWN \cite{vos80}, PZ81 \cite{per81}, 
PW92 \cite{per92}, Chachiyo \cite{Chachiyo2016J.Chem.Phys.145_021101}, revChachiyo \cite{Karasiev2016J.Chem.Phys.145_157101},
and GGA-PBE \cite{Perdew1996Phys.Rev.Lett.77_3865}.
The numerical calculation was carried out by use of \textsc{ADPACK} \cite{ADPACK}.
\par
\begin{figure}[tb]
  \centering
  \includegraphics[width=1.0\linewidth]{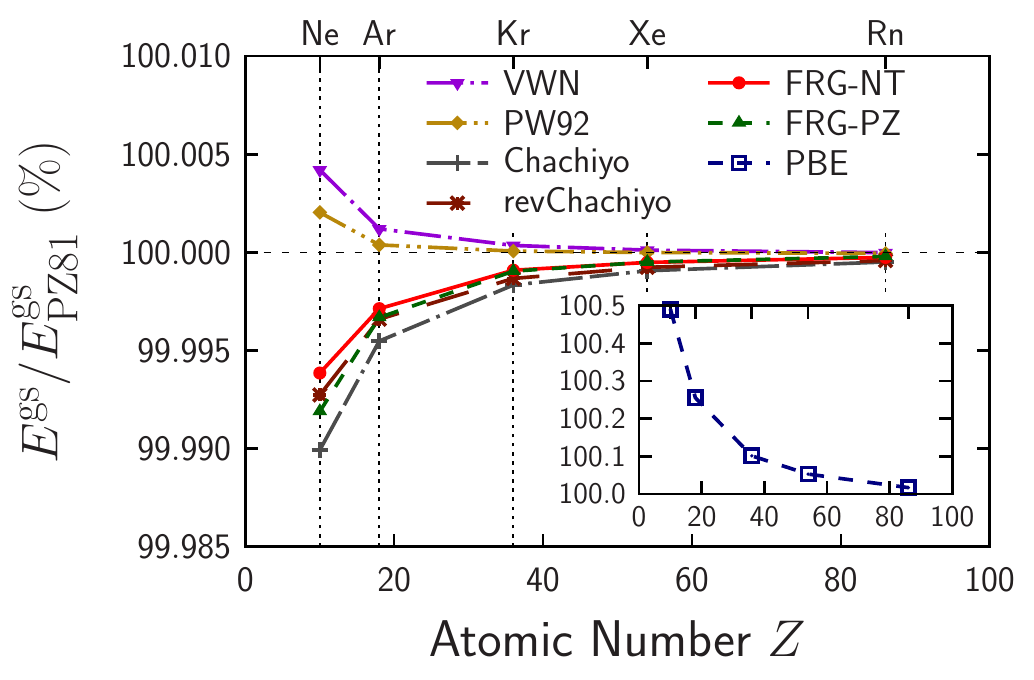}
  \caption{
    Ratios of the ground-state energies to that given by PZ81 
    shown as functions of the atomic number $ Z $.}
  \label{fig:egs}
\end{figure}

Figure~\ref{fig:egs} shows the ground-state energies 
$E^{\rm gs}$
of $ \mathrm{Ne} $, $ \mathrm{Ar} $, $ \mathrm{Kr} $, $ \mathrm{Xe} $, and $ \mathrm{Rn} $ atoms 
obtained by each EDF as ratios to the results of PZ81
$E^{\rm gs}_{\rm PZ81}$.
One can see that the functionals constructed 
from the FRG-DFT
show comparable results to those of other LDA EDFs for every atoms.
On the other hand, 
the GGA-PBE results
are quite different from those of the LDA EDFs.
This suggests that the results of our functional
reside near that of the exact LDA EDFs,
i.e.,~the LDA EDF constructed 
from the exact correlation energy per particle
$ \ec^{\urm{exact}} \left( \rs \right) $,
as much as other conventional LDA EDFs,
and the differences between our functional and other LDA EDFs
are insignificant for the accuracy in comparison with
the effect of the ignorance of the gradient.
\begin{figure}[tb]
  \centering
  \includegraphics[width=1.0\linewidth]{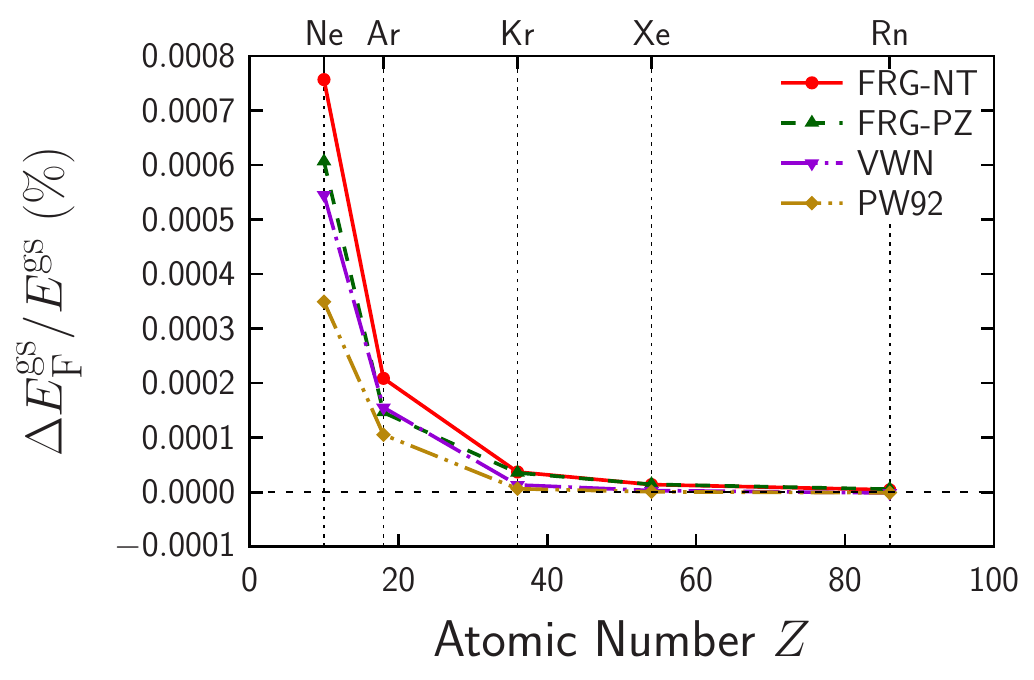}
  \caption{
    Deviations of energies from that given by PZ81 at a fixed density
    $ \rho_{\urm{gs}}^{\urm{PZ81}} \left( \ve{x} \right) $
    for each EDF shown as the ratio to the ground-state energy obtained by the EDF.}
  \label{fig:func_err}
\end{figure}
\begin{figure}[tb]
  \centering
  \includegraphics[width=1.0\linewidth]{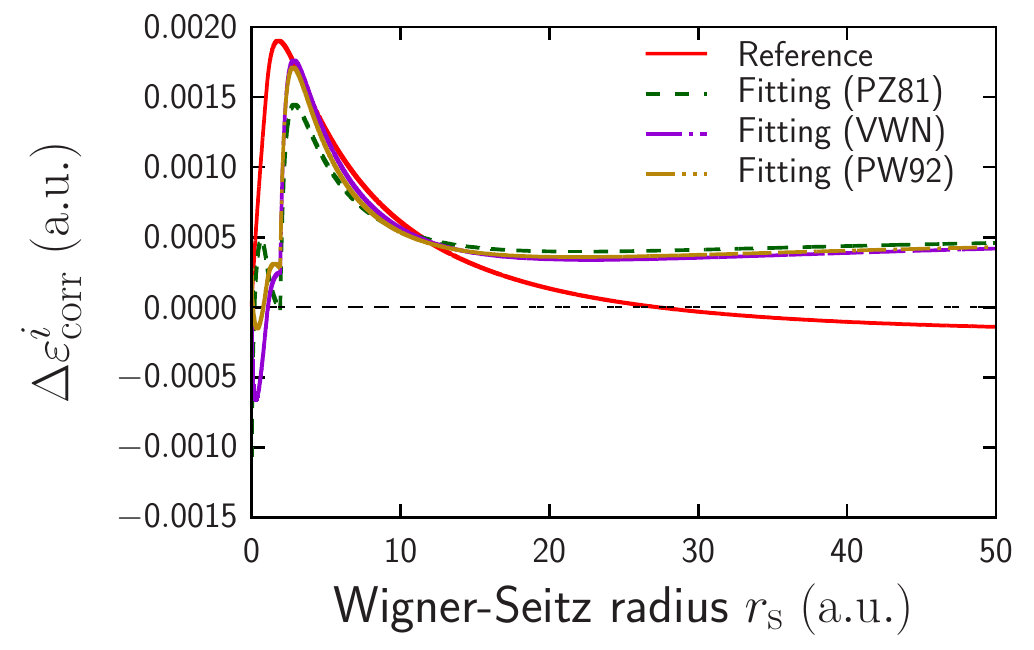}
  \caption{
    Dependence on $ \rs $ of 
    $ \decref^{\urm{FRG}} - \decref^{\urm{DMC}} $
    (Reference),
    $\decfit^{\urm{PZ81}} $ (Fitting (PZ81)),
    $\decfit^{\urm{VWN}} $ (Fitting (VWN)),
    and
    $\decfit^{\urm{PW92}} $ (Fitting (PW92)).}
  \label{fig:fit_ref_err}
\end{figure}
\par
\textit{Origins of the difference among functionals\/.}~To further understand the difference between LDA EDFs,
we investigate the difference of the energy with fixing the density
inspired by the notion of the functional-driven error \cite{Kim2013Phys.Rev.Lett.111_073003}.
Figure \ref{fig:func_err} shows 
$ \Delta E_{\urm{F}}^{\rm gs}
=
E \left[ \rho_{\urm{gs}}^{\urm{PZ81}} \right]
-
E_{\urm{PZ81}} \left[ \rho_{\urm{gs}}^{\urm{PZ81}} \right] $,
i.e.,~the deviation of each EDF $ E \left[ \rho \right] $ from PZ81 $ E_{\urm{PZ81}} \left[ \rho \right] $
at $ \rho \left( \ve{x} \right) = \rho_{\urm{gs}}^{\urm{PZ81}} \left( \ve{x} \right) $
being the ground-state density obtained by PZ81.
One can see 
that the deviations among EDFs are comparable
even at the same density.
In the case of LDA, this deviation originates from $ \ec \left( \rs \right) $,
the error of which is expected to be attributed to two parts:
the reference-driven error and the fitting-driven error, 
i.e.~the errors caused by the choice of the reference data
and the fitting functions, respectively.
By recasting the difference between EDFs in terms of these two errors,
further understanding of the origin of $ \Delta E_{\urm{F}}^{\urm{gs}} $
shown in Fig.~\ref{fig:func_err} will be obtained.
\par
We roughly assume that $ \ec^i \left( \rs \right) $ standing for $ \ec \left( \rs \right) $ 
used for functional $ i $
($ = \text{VWN} $, PZ81, PW92, FRG-NT, FRG-PZ)
is written as
\begin{align*}
\ec^i \left( \rs \right)
=
\ec^{\urm{exact}} \left( \rs \right)
+
\decref^i \left( \rs \right) 
+
\decfit^i \left( \rs \right),
\end{align*}
with the reference-driven error
$ \decref^i \left( \rs \right) $ 
and fitting-driven error
$ \decfit^i \left( \rs \right) $.
Since FRG-NT does not rely on any fitting function, 
\begin{align*}
	\decfit^{\urm{FRG-NT}} \equiv 0.
\end{align*}
The fact that the same fitting scheme is employed for FRG-PZ and PZ81
  leads to 
\begin{align*}
	\decfit^{\text{FRG-PZ}} \left( \rs \right) = \decfit^{\urm{PZ81}} \left( \rs \right).
\end{align*}
We also have 
\begin{align*}
	&
	\decref^i \left( \rs \right) 
	\\
	&= 
	\begin{cases}
	\decref^{\urm{DMC}} \left( \rs \right)
	&
	(i=\text{VWN, PZ81, PW92}),
	\\
	\decref^{\urm{FRG}} \left( \rs \right)
	&
	(i=\text{FRG-NT, FRG-PZ}),
	\end{cases}
\end{align*}
where $ \decref^{\urm{DMC}} \left( \rs \right) $ 
and $ \decref^{\urm{FRG}} \left( \rs \right) $ 
are errors stemming from the choice of DMC and FRG data, respectively.
By use of these conditions, 
$ \decfit^i \left( \rs \right) $ and
$ \decref^{\urm{FRG}} \left( \rs \right) - \decref^{\urm{DMC}} \left( \rs \right) $ 
are estimated from $ \ec^i \left( \rs \right) $.
\par
Figure~\ref{fig:fit_ref_err} shows 
$ \decfit^i \left( \rs \right) $ 
and $ \decref^{\urm{FRG}} \left( \rs \right) - \decref^{\urm{DMC}} \left( \rs \right) $.
In $ \rs \gtrsim 10 \, \mathrm{a.u.} $,
the use of fitting
affects the value of $ \ec \left( \rs \right) $
more than the choice of the 
reference data,
while these quantities have comparable magnitude
in $ \rs \lesssim 10 \, \mathrm{a.u.} $
This may explain why the comparable results for each functional are obtained in Fig.~\ref{fig:func_err}
since $ \rs \lesssim 10\, \mathrm{a.u.} $ is physically relevant for atoms.
\par
\textit{Conclusion\/.}~We presented an \textit{ab initio} construction of 
the energy density functional (EDF) for
three-dimensional electron systems using the functional-renormalization-group-aided density functional theory (FRG-DFT).
The derived correlation energies of the homogeneous electron gas
agree with the Monte-Carlo results
in a wide range of densities
reproducing the exact behavior given 
by the Gell-Mann--Brueckner resummation at the high-density limit.
Using the FRG-DFT data obtained densely for various densities,
we construct the EDF in the local density approximation (LDA)
without using any fitting function for physically relevant densities.
Applied to the KS calculation of 
the ground-state energies of the noble gas atoms,
our functional shows comparable results to
other conventional ones in LDA.
Our results show that FRG-DFT can become 
a practical method contributing to the non-empirical
construction of EDFs of realistic quantum many-body systems.
\par
Although we have focused on the case of LDA 
in this Letter, our formalism is also applicable 
for the construction of EDFs incorporating 
the effect of the gradient of density. 
There are some technical ideas 
to realize the inclusion of gradient effects, 
such as the use of the weighted density approximation or the derivative expansion, which has been developed in the context of FRG. 
Based on these ideas, we believe that the construction of EDFs beyond LDA without any empirical parameter is achievable. Our formalism and procedure can also be naturally extended to the case of constructing EDFs in the local spin density approximation, which will be presented in a forthcoming paper.
\begin{acknowledgments}
  \textit{Acknowledgments\/.}~The authors acknowledge Osamu Sugino for valuable comments on 
  the manuscript
  and
  Ryosuke Akashi, Le Minh Cristian, Haruki Kasuya, Teiji Kunihiro, Peter Maksym, Shinji Tsuneyuki, and Kenichi Yoshida
  for fruitful discussions.
  T.~Y.~was supported by the Grants-in-Aid for JSPS fellows (Grant No.~20J00644).
  T.~N.~was supported by the Grants-in-Aid for JSPS fellows (Grant No.~19J20543).
  Numerical computation in this work was carried out on Cray XC
  at the Yukawa Institute Computer Facility
  and cluster computers at the RIKEN iTHEMS program.
\end{acknowledgments}
%
%
%
\end{document}